\documentclass[aps,twocolumn,prl,showpacs,amsmath,amssymb,10pt]{revtex4-1}
\usepackage{graphicx}
\usepackage{scrextend}
\usepackage{manfnt,pifont}
\usepackage{hyperref}
\usepackage{url}
\usepackage{xcolor}
\usepackage{url}
\usepackage[normalem]{ulem}

\usepackage{CJK}

\begin{document}

\begin{CJK*}{GB}{}
\CJKfamily{gbsn}

\title{Yangian Symmetry in Holographic Correlators}

\author{Konstantinos C. Rigatos}
\affiliation{Department of Physics and Center for Theory of Quantum Matter, 390 UCB University of Colorado Boulder, CO 80309, USA}
\author{Xinan Zhou (ÖÜÏ¡éª)}
\affiliation{Kavli Institute for Theoretical Sciences, University of Chinese Academy of Sciences, Beijing 100190, China.}

\begin{abstract}
\noindent We point out that an infinite class of Witten diagrams is invariant under a Yangian symmetry. These diagrams are building blocks of holographic correlators and are related by a web of differential recursion relations. We show that Yangian invariance is equivalent to the consistency conditions of the recursion relations.

\end{abstract}

\maketitle
\end{CJK*}

\noindent{\bf Introduction.} Recently, there has been much progress in computing holographic correlators, which are the most basic observables for exploring and exploiting the AdS/CFT correspondence. For example, all four-point correlators of $\tfrac{1}{2}$-BPS operators with arbitrary Kaluza-Klein weights are known at tree level in all maximal supergravity theories \cite{Rastelli:2016nze,Rastelli:2017udc,Alday:2020lbp,Alday:2020dtb} and super Yang-Mills theory (SYM) in AdS \cite{Alday:2021odx}. Examples of higher-point correlators have also been obtained in $AdS_5$ \cite{Goncalves:2019znr,Alday:2022lkk}. While these results are highly impressive, they are all obtained by using essentially the same kind of method, namely, the bootstrap approach which imposes superconformal symmetry and physical consistency conditions \footnote{See \cite{Bissi:2022mrs} for a review of the progress in computing holographic correlators using the bootstrap method.}. It is important to ask if there are other independent guiding principles which allow us to efficiently compute holographic correlators. Particularly, in the paradigmatic example of the AdS/CFT, the 4d $\mathcal{N}=4$ SYM theory, which is dual to IIB string theory in $AdS_5\times S^5$, is known to be integrable in the planar limit. It is natural to wonder if integrability can play a role in the study of holographic correlators. Unfortunately, the standard integrability techniques are known to have difficulties in the supergravity regime \footnote{A scenario at strong coupling where integrability is tractable is correlators of heavy operators with R-symmetry weights of order $\mathcal{O}(\sqrt{N})$. See, {\it e.g.}, \cite{Jiang:2016ulr,Coronado:2018ypq,Coronado:2018cxj,Basso:2019diw,Bargheer:2019exp} for recent progress. But this is still beyond the supergravity regime where operator weights are $\mathcal{O}(1)$.}. As a result, a concrete relation between integrability and holographic correlators remains elusive. However, in this paper, we will provide hints for such a relation by pointing out that an infinite class of Witten diagrams in AdS enjoys a Yangian symmetry, which is a hallmark of integrability. While we consider only bosonic symmetry here, we hope that the analysis can be generalized to the supersymmetric case as well.

\begin{figure}[h] 
\centering
\includegraphics[width=0.30\textwidth]{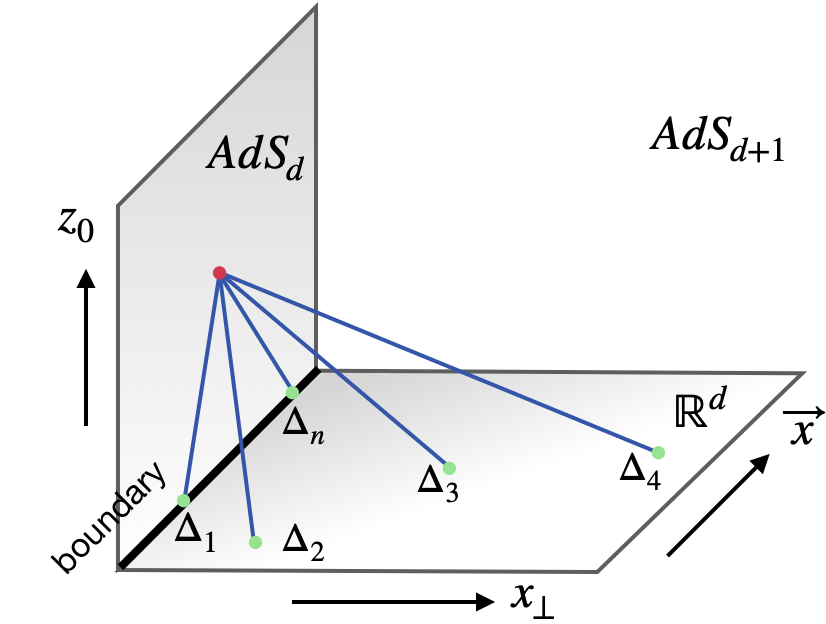}
\caption{A contact Witten diagram in Poincar\'e coordinates.}
\label{fig:wittend}
\end{figure}

More precisely, we consider the contact Witten diagrams depicted in Fig \ref{fig:wittend}, which appear naturally in holographic models of boundary CFTs. The vertical co-dimension 1 surface is the holographic dual of the boundary. When all insertions are moved to the boundary, the diagrams are fully within the $AdS_d$ subspace and reduce to  the so-called $D$-functions in the AdS/CFT literature. These contact Witten diagrams are the building blocks of holographic correlators. As we will show, these diagrams can be identified with the following conformal Feynman integral in $D$-dimensional flat space
\begin{equation}\label{defIn}
I_n=\int \frac{d^D x_0}{\prod_{j=1}^n (x_{j0}^2+m_j^2)^{\Delta_i}}\;,
\end{equation}
where $x_{ij}^\mu =x_i^\mu-x_j^\mu$, $x_{ij}^2=x_{ij}^\mu x_{ij,\mu}$ and $\sum_{i=1}^n \Delta_i=D$. The perpendicular distances $x_{i,\perp}$ are identified with the masses $m_i$. These integrals, which generalize box diagrams, are remarkably invariant under the conformal Yangian algebra. The discovery of this property was motivated by special cases of such diagrams appearing in the so-called fishnet theories which are known to be integrable \cite{Gurdogan:2015csr,Caetano:2016ydc,Chicherin:2017frs,Chicherin:2017cns}. Integrability of (\ref{defIn}) was first proven in the massless case, for integrals with  $n=4,6$. The proof was streamlined and extended to the massive case in \cite{Loebbert:2020hxk,Loebbert:2020glj} where it was shown that all such integrals are Yangian invariant. Since contact Witten diagrams are essentially $I_n$, it follows that they are Yangian invariant as well. On the other hand, we will show that the contact Witten diagrams satisfy an intricate web of differential recursion relations  shifting the weights $\Delta_i$. For example, there are differential operators $\mathbb{O}_{ij}$ which shift $\Delta_i$ and $\Delta_j$ by 1
\begin{equation}
\mathbb{O}_{ij} W\propto W\big|_{\Delta_{i,j}\to\Delta_{i,j}+1}\;,
\end{equation}
For these relations to be consistent, the action of $\mathbb{O}_{ij}\mathbb{O}_{kl}$ must be equal to that of $\mathbb{O}_{ik}\mathbb{O}_{jl}$ as they lead to the same contact Witten diagram. This imposes nontrivial constraints on $W$. Remarkably, we find that the full set of consistency conditions is precisely the Yangian invariance condition.

\vspace{0.3cm}
\noindent{\bf Yangian generators.} The Feynman integrals (\ref{defIn}) are invariant under the conformal group $SO(D,2)$ which is generated by ${\rm J}^a=\sum_{j=1}^n {\rm J}_j^a$. Here ${\rm J}_j^a$ are single-site generators acting on $x_j$
\begin{eqnarray}\label{genlevelzero}
\nonumber && {\rm P}_j^{\hat{\mu}}=-i\partial^{\hat{\mu}}_{x_j}\;,\quad\quad\quad\quad {\rm L}_j^{\hat{\mu}\hat{\nu}}=ix_j^{\hat{\mu}}\partial^{\hat{\nu}}_{x_j}-ix_j^{\hat{\nu}}\partial^{\hat{\mu}}_{x_j}\;,\\
&& {\rm D}_j=-i(x_{j,\mu}\partial^{\mu}_{x_j}+m_j\partial_{m_j}+\Delta_j)\;,\\
\nonumber && {\rm K}^{\hat{\mu}}_j=-2ix^{\hat{\mu}}_j(x_{j,\nu}\partial^{\nu}_{x_j}+m_j\partial_{m_j}+\Delta_j)+i(x_j^2+m_j^2)\partial^{\hat{\mu}}_{x_j}\;,
\end{eqnarray}
and $\mu$ runs from 1 to $D$. The index $\hat{\mu}$ runs from 1 to $D+1$, but only $\hat{\mu}=1,\ldots,D$ correspond to the symmetries of $I_n$. Note that with $\hat{\mu}=1,\ldots,D+1$, (\ref{genlevelzero}) are the $SO(D+1,2)$ conformal generators in $D+1$ dimensions where the $(D+1)$-th dimension is $x^{D+1}=m$. Conformal symmetry is partially broken along this dimension to $SO(D,2)$. The symmetry breaking is exactly the same as inserting a boundary at $x^{D+1}=0$. 

The massive Yangian is generated by the above level-zero generators and the following level-one generators \footnote{See \cite{Loebbert:2016cdm} for a pedagogical introduction.}
\begin{equation}
\widehat{{\rm J}}^a=\frac{1}{2}f^a{}_{bc}\sum_{j<k}^n {\rm J}_j^c{\rm J}_k^b+\sum_{j=1}^n s_j {\rm J}_j^a\;,
\end{equation}
where $f^a{}_{bc}$ are the structure constants and $s_i$ are the evaluation parameters. The integrals $I_n$ are annihilated by the level-one generators, and consequently the entire Yangian. Moreover, $I_n$ are invariant under level-zero generators and the level-one generators are in the adjoint representation of the level-zero algebra. It is therefore sufficient to require that $I_n$ is annihilated by the level-one momentum operators $\widehat{{\rm P}}^\mu$. Furthermore, because (\ref{defIn}) also has permutation symmetry, invariance under $\widehat{{\rm J}}^a$ is equivalent to invariance under any two-site operators \cite{Loebbert:2020glj}
\begin{equation}
\widehat{{\rm J}}^a_{jk}=\frac{1}{2}f^a{}_{bc}{\rm J}_j^c{\rm J}_k^b+\frac{\Delta_k}{2} {\rm J}_j^a-\frac{\Delta_j}{2} {\rm J}_k^a\;.
\end{equation}
In terms of $\widehat{{\rm J}}^a_{jk}$, $\widehat{{\rm J}}^a$ can be written as $\widehat{{\rm J}}^a=\sum_{k>j=1}^n\widehat{{\rm J}}^a_{jk}$. The momentum operator is given by 
\begin{equation}
\widehat{\rm P}^\mu_{jk}=\frac{i}{2}\left({\rm P}^\mu_j{\rm D}_k+{\rm P}_{j,\nu}{\rm L}^{\mu\nu}_k-i\Delta_k {\rm P}^\mu_j-(j\leftrightarrow k)\right)\;.
\end{equation}
Using (\ref{genlevelzero}), we can write it explicitly as 
\begin{equation}\label{Phatdiff}
\begin{split}
\widehat{\rm P}^\mu_{jk}={}&\frac{i}{2}\big(X^{\nu\mu\rho}\partial_{x_j,\rho}\partial_{x_k,\nu}+(2\Delta_j+m_j\partial_{m_j})\partial^\mu_{x_k}\\
{}&-(2\Delta_k+m_k\partial_{m_k})\partial^\mu_{x_j}\big)\;,
\end{split}
\end{equation}
where 
\begin{equation}
X^{\nu\mu\rho}=x_{jk}^\nu \eta^{\mu\rho}+x_{jk}^\rho \eta^{\mu\nu}-x_{jk}^\mu \eta^{\nu\rho}\;.
\end{equation}
In addition to the above operators $\widehat{{\rm J}}^a_{jk}$, it was observed in \cite{Loebbert:2020hxk} that $I_n$ are also annihilated by an extra set of bi-local operators $\widehat{{\rm J}}^a_{{\rm extra}, jk}$. For example, 
\begin{equation}
\begin{split}
\widehat{{\rm P}}^\mu_{jk,{\rm extra}}={}&\frac{i}{2}\big({\rm P}_{j,D+1}{\rm L}_k^{\mu, D+1}-(j\leftrightarrow k)\big)\\
={}&\frac{i}{2}\left(\partial_{m_j}x^\mu_k \partial_{m_k}-\partial_{m_j}m_k \partial_{x_k}^\mu -(j\leftrightarrow k)\right)\;.
\end{split}
\end{equation}
Here, we have written down a mass $m_i$ for each site $i$. The massless (or partially massless) case is obtained by just setting the masses to zero.

\vspace{0.3cm}
\noindent{\bf Witten diagrams.} The contact Witten diagram in Fig. \ref{fig:wittend} is defined as an integral over $AdS_d$
\begin{equation}\label{defW}
W=\int \frac{dz_0 d^{d-1}\vec{z}}{z_0^d}\prod_{i=1}^nG^{\Delta_i}_{B\partial}(z,\vec{x}_i,m_i)\;,
\end{equation}
where $G^{\Delta_i}_{B\partial}$ are the bulk-to-boundary propagators
\begin{equation}
G^{\Delta_i}_{B\partial}(z,\vec{x}_i,m_i)=\bigg(\frac{z_0}{z_0^2+(\vec{z}-\vec{x}_i)^2+m_i^2}\bigg)^{\Delta_i}\;.
\end{equation}
These diagrams arise in holographic models of boundary CFTs or interface CFTs where the defect is a probe brane \cite{Karch:2001cw,Karch:2000gx,DeWolfe:2001pq,Aharony:2003qf,Rastelli:2017ecj}. They are generated by contact vertices which are localized on the $AdS_d$ subspace. When all masses are zero, $W$ reduces to the $D$-function $D_{\Delta_1,\ldots,\Delta_n}$ in $AdS_d$. Note that unlike the Feynman integral $I_n$, there is no constraint relating $\Delta_i$ and $d$. The conformal invariance of $W$ is inherited from the isometry of AdS. These contact diagrams have been systematically studied in \cite{Rastelli:2017ecj} and we will use its results to establish the equivalence between $W$ and $I_n$. 

A particularly useful representation of $W$ is \cite{Rastelli:2017ecj} 
\begin{equation}\label{WtPm}
W=C_n\int_0^\infty\prod_{i=1}^n dt_i t_i^{\Delta_i-1}e^{-\sum_{i<j}t_i t_j P_{ij}-(\sum_{i=1}^n t_i m_i)^2}\;,
\end{equation}
which is obtained by using the Schwinger parameterization and integrating out the AdS coordinates. Here, $C_n=\pi^{\frac{d-1}{2}}\Gamma[\frac{\sum_{i=1}^n\Delta_i-d+1}{2}]\prod_{i=1}^n\Gamma^{-1}[\Delta_i]$ is a coefficient and we have defined 
\begin{equation}
P_{ij}=x_{ij}^2+(m_i-m_j)^2\;.
\end{equation}
An important consequence of (\ref{WtPm}) is that contact Witten diagrams are {\it dimension-independent} after factoring out a numerical coefficient
\begin{equation}
\widetilde{W}=C_n^{-1}W\;.
\end{equation}
On the other hand, if we integrate out only the radial coordinate $z_0$, we find
\begin{equation}\label{WtSchwinger}
\begin{split}
\widetilde{W}={}&\frac{\pi^{\frac{1-d}{2}}}{2}\int d^{d-1}\vec{z}\int_0^\infty \prod_{i=1}^n dt_i t_i^{\Delta_i-1}\\{}&\times \big(\sum_{i=1}^n t_i\big)^{\frac{d-1-\sum_{i=1}^n\Delta_i}{2}}
e^{-\sum_{i=1}^n t_i((\vec{z}-\vec{x}_i)^2+m_i^2)}\;.
\end{split}
\end{equation}
Using the $d$-independence of $\widetilde{W}$, we can conveniently set $d=D+1$. Then (\ref{WtSchwinger}) is nothing but the conformal integral $I_n$ after using the Schwinger parameterization
\begin{equation}
\widetilde{W}=\frac{\pi^{-\frac{\sum_{i=1}^n\Delta_i}{2}}\prod_{i=1}^n\Gamma[\Delta_i]}{2}I_n\;.
\end{equation}
Since the integrals $I_n$ are invariant under the Yangian \cite{Loebbert:2019vcj,Loebbert:2020hxk,Loebbert:2020glj}, the contact Witten diagrams $W$ are Yangian invariant as well.

\vspace{0.3cm}
\noindent{\bf Recursions and consistency conditions.} The representation (\ref{WtPm}) also makes the recursion relations of Witten diagrams manifest. Let us denote 
\begin{equation}
\mathbb{O}_{ij}=\frac{\partial}{\partial P_{ij}}\big|_{P,m}\;,\quad\quad N_i=\frac{\partial}{\partial m_i}\big|_{P,m}
\end{equation}
as the partial derivatives where $P_{ij}$, $m_i$ are regarded as the independent variables. Then $N_i$ is related to $\partial_{m_i}$ in (\ref{genlevelzero}), where $x_i^\mu$ and $m_i$ are regarded as the independent variables, by 
\begin{equation}\label{mpartialm}
m_i\partial_{m_i}=\mathbb{D}_i+2\sum_{j\neq i}m_i^2 \mathbb{O}_{ij}\;,
\end{equation}
and we have defined 
\begin{equation}
\mathbb{D}_i=m_iN_i-2\sum_{j\neq i}m_im_j \mathbb{O}_{ij}\;.
\end{equation}
From the integral representation (\ref{WtPm}), it is obvious that we have the following {\it differential recursion relations}
\begin{eqnarray}
\mathbb{O}_{ij}W&=&\frac{2\Delta_i\Delta_j}{d-1-\sum_i \Delta_i} W\big|_{\Delta_{i,j}\to \Delta_{i,j}+1}\;,\label{diffrec1}\\
\mathbb{D}_iW&=&\frac{4m_i^2\Delta_i(\Delta_i+1)}{d-1-\sum_i \Delta_i} W\big|_{\Delta_i\to \Delta_i+2}\label{diffrec2}\;,
\end{eqnarray}
which shift the conformal dimensions. These relations generalize the well known weight-shifting relations of $D$-functions \cite{DHoker:1999kzh}. However, the relations (\ref{diffrec1}) and (\ref{diffrec2}) must give the same answer when reaching the same point in weight space following different paths. This gives rise to the following {\it consistency conditions}
\begin{eqnarray}
&(\mathbb{O}_{ij}\mathbb{O}_{kl}-\mathbb{O}_{ik}\mathbb{O}_{jl})W=0\;, \quad i,l\neq j,k\;,&\label{massiveconsistency1}\\
&\mathbb{D}_i\mathbb{O}_{kl}W=2m_i^2 \mathbb{O}_{ik}\mathbb{O}_{il}W\;,\quad i,j,k\text{ all different}\;,&\label{massiveconsistency2}\\
&\mathbb{D}_j\mathbb{D}_kW=4m_j^2m_k^2 \mathbb{O}_{jk}\mathbb{O}_{jk}W\;,\quad\quad j\neq k\;.&\label{massiveconsistency3}
\end{eqnarray}
Note that these conditions are also satisfied by $I_n$ because they are identical to $W$ up to overall coefficients. 

Let us also mention that the conformal invariance of Witten diagrams implies the following relations
\begin{equation}\label{massiveconfcov}
(m_iN_i+P_{ij}\mathbb{O}_{ij})W=-\Delta_i W\;.
\end{equation}
These conditions can be easily derived in the embedding space formalism and they follow from requiring $W$ to scale correctly when independently rescaling the embedding vector of each operator  \cite{Rastelli:2017ecj}. The details can be found in the Supplemental Material.

\vspace{0.3cm}
\noindent{\bf Yangian constraints as consistency conditions.} We now show that the Yangian invariance conditions
\begin{eqnarray}
\widehat{{\rm P}}^\mu_{jk} W &=&0\;,\label{condiPhat}\\
\widehat{{\rm P}}^\mu_{jk,{\rm extra}} W&=&0\label{condiPhatextra}\;,
\end{eqnarray}
are equivalent to the consistency conditions of the recursion relations (\ref{massiveconsistency1}), (\ref{massiveconsistency2}), (\ref{massiveconsistency3}). Instead of working with cross ratios, which spoils  manifest permutation symmetry, we work with the variables $P_{ij}$ and $m_i$. Then using 
\begin{equation}
\begin{split}
{}&\partial_{x_j}^\mu=2\sum_{i\neq j} x_{ji}^\mu \mathbb{O}_{ij}\;,\\
{}&\partial_{x_j}^\rho \partial_{x_k}^\nu=4\sum_{i\neq k}\sum_{l\neq j}x_{jl}^\rho x_{ki}^\nu \mathbb{O}_{jl}\mathbb{O}_{ki}-2\eta^{\rho\nu}\mathbb{O}_{jk}\;,
\end{split}
\end{equation}
and (\ref{mpartialm}) we can take all derivatives with respect to $P_{ij}$ and $m_i$. We will find that the action of the operators can be written in the form
\begin{equation}\label{PTE}
\begin{split}
{}&-2i\widehat{{\rm P}}^\mu_{jk} W = \sum_{a<b} {\rm T}_{ab}^\mu E_{ab}\;,\\
 {}&-2i\widehat{{\rm P}}^\mu_{jk,{\rm extra}} W = \sum_{a<b} {\rm T}_{ab}^\mu E_{ab,{\rm extra}}\;,
\end{split}
\end{equation}
where ${\rm T}_{ab}^\mu=\frac{x_{ab}^\mu}{P_{ab}}$. The coefficients $E_{ab}$, $E_{ab,{\rm extra}}$ have the same scaling dimensions as $W$. It was shown in \cite{Loebbert:2020glj} that ${\rm T}_{ab}^\mu$ are linearly independent with respect to coefficients which are functions of cross ratios \footnote{Here we are assuming the spacetime dimension $D$ is high enough with respect to $n$.}. Yangian invariance then requires that all coefficient functions $E_{ab}$, $E_{ab,{\rm extra}}$ must vanish separately. The upshot is that these conditions boil down to the three basic relations (\ref{massiveconsistency1}), (\ref{massiveconsistency2}) and (\ref{massiveconsistency3}).

\vspace{0.3cm}
{\it The massless case.} For simplicity, let us first demonstrate the equivalence for the massless case, {\it i.e.}, $m_i=0$, which is relevant for $D$-functions in pure AdS. Note that $\widehat{{\rm P}}^\mu_{jk,{\rm extra}}$ vanishes in this case so we have only (\ref{condiPhat}) with $m_i$ set to zero. From (\ref{Phatdiff}) it is not difficult to see that almost all terms are already in the form of (\ref{PTE}), except for those coming from the contraction with $X^{\nu\mu\rho}$. To proceed, we note the following useful identity
\begin{equation}\label{Xxx}
\begin{split}
{}&X^{\nu\mu\rho} x_{jl}^\rho x_{ki}^\nu=\frac{1}{2}\big({\rm T}_{jk}^\mu P_{jk}P_{li}-{\rm T}_{ji}^\mu P_{ji}P_{kl}-{\rm T}_{jl}^\mu P_{jl}P_{ki}\\
{}&\quad\quad+{\rm T}_{ki}^\mu P_{ki}P_{jl}+{\rm T}_{kl}^\mu P_{kl}P_{ij}-{\rm T}_{il}^\mu P_{il}P_{jk}\big)\;.
\end{split}
\end{equation}
We then find all the coefficient functions are given by 
\begin{eqnarray}
&&E_{il}=-2P_{il}P_{jk}(\mathbb{O}_{jl}\mathbb{O}_{ik}-\mathbb{O}_{ji}\mathbb{O}_{kl})W\;,\label{condiEila}\\
\nonumber && E_{ki}=2\big\{\sum_{l\neq j,k} P_{ki}P_{jl}\mathbb{O}_{jl}\mathbb{O}_{ki}+2P_{ki}P_{jk}\mathbb{O}_{jk}\mathbb{O}_{ki}\\
&&\quad\quad\quad+\sum_{l\neq j,k}P_{ki}P_{jl}\mathbb{O}_{ji}\mathbb{O}_{kl}+2\Delta_j P_{ki}\mathbb{O}_{ki}\big\}W\;,\label{condiEika}\\
&& E_{jl}= -E_{ki}\big|_{j\leftrightarrow k, i\leftrightarrow l}\;,\label{condiEjla}\\
\nonumber &&E_{jk}=2\big\{ \sum_{i,l\neq j,k} P_{jk}P_{il}\mathbb{O}_{jl}\mathbb{O}_{ki}-2P_{jk}^2\mathbb{O}_{jk}\mathbb{O}_{jk}\\
&&\quad\quad\quad-(2-D+2\Delta_j+2\Delta_k)P_{jk}\mathbb{O}_{jk}\big\}W\;,\label{condiEjka}
\end{eqnarray}
where $i,l\neq j,k$ \footnote{The indices $i$ and $l$ are dummy indices. Therefore, $E_{kl}$ and $E_{ji}$ are not new coefficient functions.}. From $E_{il}=0$, we reproduce the consistency condition (\ref{massiveconsistency1}). Since (\ref{massiveconsistency2}) and (\ref{massiveconsistency3}) are identically zero on both sides in the massless limit, the remaining conditions must not produce nontrivial constraints. To show $E_{ki}=0$, we first use (\ref{massiveconsistency1}) to write $E_{ki}$ as
\begin{equation}
\begin{split}
E_{ki}={}&4\big\{\sum_{l\neq j,k} P_{ki}P_{jl}\mathbb{O}_{jl}\mathbb{O}_{ki}+P_{ki}P_{jk}\mathbb{O}_{jk}\mathbb{O}_{ki}\\
{}&+\Delta_j P_{ki}\mathbb{O}_{ki}\big\}W\;.
\end{split}
\end{equation}
Then using the massless limit of (\ref{massiveconfcov}) we find that $E_{ki}$ vanishes. Symmetry implies that $E_{jl}=0$ as well. To see $E_{jk}=0$, we use permutation symmetry and (\ref{massiveconsistency1}) to write 
\begin{equation}
\nonumber \sum_{i,l\neq j,k} P_{jk}P_{il}\mathbb{O}_{jl}\mathbb{O}_{ki}W=\sum_{i,l\neq j,k} P_{jk}P_{il}\mathbb{O}_{jk}\mathbb{O}_{il}W\;.
\end{equation}
From (\ref{massiveconfcov}) we also have
\begin{equation}
\nonumber \sum_{i,l\neq j,k} P_{il}\mathbb{O}_{il}W=(-D+2\Delta_j+2\Delta_k+2P_{jk}\mathbb{O}_{jk})W\;,
\end{equation}
where we have used $D=\sum_{i=1}^n\Delta_i$. It is then clear that $E_{jk}$ also vanishes. 

\vspace{0.3cm}
{\it The massive case.} Having proven the equivalence in the massless limit, let us now move on to the general case. We first focus on the condition (\ref{condiPhat}) where the proof is similar to the massless case above. To cast the action of $\widehat{{\rm P}}^\mu_{jk}$ in the form of (\ref{PTE}), let us use the following massive version of (\ref{Xxx})
\begin{equation}\label{Xxxmassive}
\begin{split}
\nonumber  {}&X^{\nu\mu\rho} x_{jl}^\rho x_{ki}^\nu=\frac{1}{2}\big({\rm T}_{jk}^\mu P_{jk}P_{li}-{\rm T}_{ji}^\mu P_{ji}P_{kl}-{\rm T}_{jl}^\mu P_{jl}P_{ki}\\
{}&\quad\quad+{\rm T}_{ki}^\mu P_{ki}P_{jl}+{\rm T}_{kl}^\mu P_{kl}P_{ij}-{\rm T}_{il}^\mu P_{il}P_{jk}\big)\\
{}&\quad\quad+x_{jl}^\mu m_k(m_k-m_i)-x_{ki}^\mu m_j(m_j-m_l)\\
{}&\quad\quad+x_{ij}^\mu m_km_l+x_{jk}^\mu m_im_l+x_{kl}^\mu m_im_j+x_{li}^\mu m_j m_k\;.
\end{split}
\end{equation}
We find the coefficient functions are
\begin{eqnarray}
&&P_{il}^{-1}E_{il}=-2(P_{jk}+2m_jm_k)(\mathbb{O}_{jl}\mathbb{O}_{ki}-\mathbb{O}_{ji}\mathbb{O}_{kl})W\;, \label{condiEilb}\\
\nonumber &&P_{ki}^{-1}E_{ki}=2\big\{(2\Delta_j+m_jN_j)\mathbb{O}_{ki}+2m_jm_k\mathbb{O}_{jk}\mathbb{O}_{ki}\\
\nonumber &&\quad\quad +\sum_{l\neq k}(P_{jl}+2m_lm_j)\mathbb{O}_{ji}\mathbb{O}_{kl}+\sum_{l\neq j}P_{jl} \mathbb{O}_{jl}\mathbb{O}_{ki}\\
&&\quad\quad+P_{jk}\mathbb{O}_{jk}\mathbb{O}_{ki}\big\}W\;, \label{condiEkib}\\
&& P_{jl}^{-1}E_{jl}=-P_{ki}^{-1}E_{ki}\big|_{j\leftrightarrow k, i\leftrightarrow l}\;, \label{condiEjlb}\\
\nonumber && P_{jk}^{-1}E_{jk}=2\big\{ \sum_{i,l\neq j,k} P_{il}\mathbb{O}_{jl}\mathbb{O}_{ki}-2P_{jk}\mathbb{O}_{jk}\mathbb{O}_{jk}\\
\nonumber &&\quad\quad+(D-2-2\Delta_j-2\Delta_k-m_jN_j-m_kN_k)\mathbb{O}_{jk}\\
&&\quad\quad+2\sum_{i\neq k}\sum_{l\neq j}m_im_l\mathbb{O}_{jl}\mathbb{O}_{ki}-2m_jm_k \mathbb{O}_{jk}\mathbb{O}_{jk}\big\}W\;, \label{condiEjkb}
\end{eqnarray}
with $i,l\neq j,k$. Requiring (\ref{condiEilb}) to vanish, we recover the first consistency condition (\ref{massiveconsistency1}). Following similar manipulations as in the massless case, which are detailed in the Supplemental Material, we find from $E_{ki}=0$ the second consistency condition (\ref{massiveconsistency2}). However, the condition from the coefficient (\ref{condiEjkb}) yields no further constraint. In fact, we find that $E_{jk}=0$ follows from (\ref{massiveconsistency1}) and (\ref{massiveconsistency2}). To derive the last consistency condition (\ref{massiveconsistency3}), we must examine the extra contraint (\ref{condiPhatextra}). The explicit operator action reads
\begin{equation}\label{Phatextraexp}
\begin{split}
{}&-2i m_j m_k\widehat{{\rm P}}^\mu_{jk,{\rm extra}} W = -x_{jk}^\mu m_j\partial_{m_j} m_k\partial_{m_k}W\\
{}& -2\sum_{l\neq k}x_{kl}^\mu \mathbb{O}_{kl}m_k^2 m_j\partial_{m_j}W+2\sum_{i\neq j}x_{ji}^\mu \mathbb{O}_{ji}m_j^2 m_k\partial_{m_k}W\;,
\end{split}
\end{equation}
where $m_j\partial_{m_j}$ should be expressed in terms of $\mathbb{D}_j$ and $\mathbb{O}_{ij}$ using (\ref{mpartialm}). Naively,  the form of (\ref{Phatextraexp}) seems to be in contradiction with (\ref{PTE}). However, this expression can be greatly simplified upon using  (\ref{massiveconsistency1}), (\ref{massiveconsistency2}) and the conformal invariance condition (\ref{massiveconfcov}) (see Supplemental Material). We find that all $E_{ab,{\rm extra}}$ vanish except for $E_{jk,{\rm extra}}$
\begin{equation}
E_{jk,{\rm extra}} = -P_{jk} (\mathbb{D}_j\mathbb{D}_k-4m_j^2m_k^2\mathbb{O}_{jk}\mathbb{O}_{jk})W\;,
\end{equation}
which gives the last condition (\ref{massiveconsistency3}).

\vspace{0.3cm}
\noindent{\bf Discussions and outlook.} In this paper, we established a new connection between integrability and holography by reinterpreting Yangian invariant Feynman integrals as Witten diagrams in AdS. We also provided an interesting reformulation of the Yangian constraints as the consistency conditions of weight-shifting relations satisfied by Witten diagrams. These conditions are obtained explicitly as (\ref{massiveconsistency1}), (\ref{massiveconsistency2}), (\ref{massiveconsistency3}), and are valid for arbitrary $n$-point functions. Compared to the original Yangian invariance constraints (\ref{condiPhat}) and (\ref{condiPhatextra}), these conditions no longer contain redundancies and are much simpler to exploit ({\it e.g.}, to explicitly compute $I_n$ as power series \cite{Loebbert:2020hxk,Loebbert:2020glj}). The remarkable simplicity of these conditions might also provide further insight into their underlying structures and hopefully open a door to applying the full power of integrability methods to holographic correlators. 

There are plenty of future directions worth exploring. Firstly, we only focused on contact Witten diagrams which correspond to one-loop Feynman integrals. It would be interesting to study Yangian symmetry in exchange Witten diagrams. Certain two-loop Feynman integrals are also known to be Yangian invariant \cite{Loebbert:2019vcj,Loebbert:2020hxk,Loebbert:2020glj} and coincide with exchange Witten diagrams when conformal dimensions satisfy special conditions \cite{Paulos:2012nu,Ma:2022ihn}. However, the general story is still unclear at the moment. Secondly, another exciting research avenue is to extend the analysis to include supersymmetry. The superconformal Yangian constraints should be highly nontrivial and will presumably select ``superspace $D$-functions'' with quantized dimensions as their solutions. It would be extremely interesting to see if these superconformal Yangian constraints can be used as an alternative method to rederive the general results of holographic four-point correlators of IIB supergravity in $AdS_5\times S^5$ \cite{Rastelli:2016nze,Rastelli:2017udc} . Finally, Witten diagrams also play an important role in the analytic functional approach to the conformal bootstrap where they serve as generating functions for the analytic functionals \cite{Mazac:2018mdx,Mazac:2018ycv,Kaviraj:2018tfd,Mazac:2018biw,Mazac:2019shk,Caron-Huot:2020adz,Giombi:2020xah}. It would be interesting to explore the consequence of Yangian symmetry in that context. 

\vspace{0.5cm} 
\noindent{\bf Acknowledgements.} We thank Yunfeng Jiang for helpful comments on the draft. The work of X.Z. is supported by funds from University of Chinese Academy of Sciences (UCAS), funds from the Kavli Institute for Theoretical Sciences (KITS), and also by the Fundamental Research Funds for the Central Universities. K.C.R.s work is supported in part by the U.S. Department of Energy (DOE), Office of Science, Office of High Energy Physics, under Award Number DE-SC0010005.

\vspace{1cm}

\appendix
\section{{\large \sc{Supplemental Material}}}
\subsection{Embedding space and the conformal invariance condition}
The action of conformal group can be linearized by going into the embedding space which has two extra dimensions. Each point $(\vec{x},m)$ in $\mathbb{R}^{D,1}$ can be represented as a null ray $P$ in $\mathbb{R}^{D+1,2}$ satisfying 
\begin{equation}
P\cdot P=0\;,\quad P\sim \lambda P\;.
\end{equation}
The conformal group $SO(D+1,2)$ acts as rotations on $P$. Operators are defined on these rays with the scaling property
\begin{equation}\label{Oscaling}
\mathcal{O}(\lambda P)=\lambda^{-\Delta} \mathcal{O}(P)\;.
\end{equation}
Explicitly, we can gauge fix the rescaling degree of freedom and parameterize the null ray as
\begin{equation}
P=\big(\frac{1+\vec{x}^2+m^2}{2},\frac{1+\vec{x}^2-m^2}{2},\vec{x},m\big)\;.
\end{equation}
Here the first two components are the two extra dimensions and have signature $-$ and $+$ respectively. With two embedding vectors, we can write down an invariant 
\begin{equation}
-2P_i\cdot P_j=x_{ij}^2+(m_i-m_j)^2=P_{ij}\;.
\end{equation}
In our case, the $SO(D+1,2)$ conformal symmetry is further broken to $SO(D,2)$. This is achieved by introducing a fixed embedding vector
\begin{equation}
B=(0,0,\vec{0},1)\;.
\end{equation} 
In addition to $P_{ij}$, we can write down another $SO(D,2)$ invariant $m_i=P_i\cdot B$. Correlators are functions of these invariants
\begin{equation}
\langle \mathcal{O}_1(P_1)\ldots \mathcal{O}_n(P_n)\rangle=W(P_{ab},m_a)\;.
\end{equation}
On the other hand, they must obey the rescaling (\ref{Oscaling}). If we let $P_i\to \lambda P_i$ then 
\begin{equation}
W(\ldots, \lambda P_{ij},\ldots,\lambda m_i,\ldots)=\lambda^{-\Delta_i}W(P_{ab},m_a)\;.
\end{equation}
This gives
\begin{equation}
(m_iN_i+P_{ij}\mathbb{O}_{ij})W=-\Delta_i W\;.
\end{equation}

\subsection{More details of the proof}
Here we give more details for the intermediate steps in the proof. To extract the consistency condition from $E_{ki}$ of the massive case, we use (\ref{massiveconsistency1}) to write it as 
\begin{equation}
\begin{split}
{}&P_{ki}^{-1}E_{ki}=2\big\{(2\Delta_j+m_jN_j)\mathbb{O}_{ki}+2m_jm_k\mathbb{O}_{jk}\mathbb{O}_{ki}\\
{}&\quad\quad +\sum_{l\neq k}2m_lm_j\mathbb{O}_{ji}\mathbb{O}_{kl}+2\sum_{l\neq j}P_{jl} \mathbb{O}_{jl}\mathbb{O}_{ki}\big\}W\;,
\end{split}
\end{equation}
Then using (\ref{massiveconfcov}) we can rewrite it as
\begin{equation}
\begin{split}
P_{ki}^{-1}E_{ki}={}&2\big\{-m_jN_j\mathbb{O}_{ki}+2m_jm_k\mathbb{O}_{jk}\mathbb{O}_{ki}\\
{}&+\sum_{l\neq k}2m_lm_j\mathbb{O}_{ji}\mathbb{O}_{kl}\big\}W\;.
\end{split}
\end{equation}
Upon using (\ref{massiveconsistency1}) again, we have 
\begin{equation}
\begin{split}
P_{ki}^{-1}E_{ki}={}&2\big\{-m_jN_j\mathbb{O}_{ki}+\sum_{l\neq j}2m_lm_j\mathbb{O}_{jl}\mathbb{O}_{ki}\\
{}&+2m_j^2\mathbb{O}_{ji}\mathbb{O}_{kj}\big\}W\;,
\end{split}
\end{equation}
which gives the second consistency condition (\ref{massiveconsistency2}). 

To prove $E_{jk}=0$, let us note $\sum_{i,l\neq j,k} P_{il}\mathbb{O}_{jl}\mathbb{O}_{ki}W=\sum_{i,l\neq j,k} P_{il}\mathbb{O}_{il}\mathbb{O}_{jk}W$ and 
\begin{equation}
\begin{split}
\nonumber \sum_{i,l\neq j,k} P_{il}\mathbb{O}_{il}\mathbb{O}_{jk}W={}&\mathbb{O}_{jk}\big\{-\sum_{i\neq j,k}m_iN_i+m_jN_j+m_kN_k\\
{}&-D+2\Delta_j+2\Delta_k+2P_{jk}\mathbb{O}_{jk}\big\}W\;,
\end{split}
\end{equation}
which follows from (\ref{massiveconfcov}). This allows us to write 
\begin{equation}
\begin{split}
\nonumber P_{jk}^{-1}E_{jk}={}&2\big\{-\sum_{i\neq j,k}m_iN_i\mathbb{O}_{jk}+2\sum_{i\neq k}\sum_{l\neq j}m_im_l\mathbb{O}_{jl}\mathbb{O}_{ki}\\
{}&-2m_jm_k\mathbb{O}_{jk}\mathbb{O}_{jk}\big\}W\;.
\end{split}
\end{equation}
From (\ref{massiveconsistency2}) we have
\begin{equation}
\begin{split}
{}&\big\{\sum_{i\neq j,k}m_iN_i\mathbb{O}_{jk}-2\sum_{i\neq j,k}\sum_{l\neq i}m_im_l\mathbb{O}_{il}\mathbb{O}_{jk}\\
{}&\quad\quad\quad\quad\quad\quad-2\sum_{i\neq j,k}m_i^2\mathbb{O}_{ij}\mathbb{O}_{ik}\big\}W=0\;.
\end{split}
\end{equation}
Using this identity and (\ref{massiveconsistency1}), we find $E_{jk}=0$.

Finally, let us look at the condition from acting with $\widehat{{\rm P}}^\mu_{jk,{\rm extra}}$. We note the following two identities 
\begin{equation}
\begin{split}
{}&\big(\sum_{i\neq j,k}x_{ji}^\mu \mathbb{O}_{ji}m_j^2 m_k\partial_{m_k}-\sum_{l\neq j,k}x_{kl}^\mu \mathbb{O}_{kl}m_k^2 m_j\partial_{m_j}\big)W\\
{}&=-2\sum_{l\neq j,k}x_{kl}^\mu m_j^2m_k^2\big(\sum_{i\neq j}\mathbb{O}_{ji}\mathbb{O}_{kl}+\mathbb{O}_{jk}\mathbb{O}_{jl}\big)W\\
{}&\quad+2\sum_{i\neq j,k}x_{ji}^\mu m_j^2m_k^2 \big( \sum_{l\neq k}\mathbb{O}_{kl}\mathbb{O}_{ji}+\mathbb{O}_{jk}\mathbb{O}_{ki}\big)W\\
{}&=2m_j^2m_k^2 x_{jk}^\mu\big(\sum_{l\neq k}\sum_{i\neq j}\mathbb{O}_{ji}\mathbb{O}_{kl}-\mathbb{O}_{jk}\mathbb{O}_{jk}\big)W\;,
\end{split}
\end{equation}
\begin{equation}
\begin{split}
{}&\big(x_{jk}^\mu m_j^2 \mathbb{O}_{jk}m_k\partial_k-x_{jk}^\mu  m_j^2 \sum_{i\neq j}\mathbb{O}_{ji}\mathbb{D}_k\big)W\\
{}&=x_{jk}^\mu\big(- m_j^2 \sum_{i\neq j,k} \mathbb{O}_{ji}\mathbb{D}_k+2m_j^2m_k^2\mathbb{O}_{jk}\sum_{l\neq k}\mathbb{O}_{kl}\big)W\\
{}&=2 x_{jk}^\mu m_j^2m_k^2\mathbb{O}_{jk}\mathbb{O}_{jk}W\;.
\end{split}
\end{equation}
In obtaining these identities we have used the definition (\ref{mpartialm}) for $m_j\partial_{m_j}$ and the first two consistency conditions (\ref{massiveconsistency1}), (\ref{massiveconsistency2}). We now expand (\ref{Phatextraexp}) and use these identities. It is straightforward to find that (\ref{Phatextraexp}) is proportional to $x_{jk}^\mu$ and reads
\begin{equation}
\begin{split}
\nonumber {}&-2i m_j m_k\widehat{{\rm P}}^\mu_{jk,{\rm extra}} W = -x_{jk}^\mu (\mathbb{D}_j\mathbb{D}_k-4m_j^2m_k^2\mathbb{O}_{jk}\mathbb{O}_{jk})W\;.
\end{split}
\end{equation}
This gives the final consistency condition (\ref{massiveconsistency3}).

\bibliography{refsYangian} 
\bibliographystyle{utphys}
\end{document}